\documentclass[11pt,a4paper,reqno]{amsart} 

\usepackage{amssymb,graphicx}

\usepackage{amssymb}

\newcommand{\koniec}{\begin{flushright}  $\Box $ \end{flushright}}
\newtheorem{theo}{Theorem}[section] 
\newtheorem{prop}[theo]{Proposition}

\newcounter{mnotecount}[section]

\renewcommand{\themnotecount}{\thesection.\arabic{mnotecount}}

\newcommand{\mnote}[1]
{\protect{\stepcounter{mnotecount}}$^{\mbox{\footnotesize
$
\bullet$\themnotecount}}$ \marginpar{
\raggedright\tiny\em
$\!\!\!\!\!\!\,\bullet$\themnotecount: #1} }

\def\be{\begin{equation}}
\def\l{\ell}
\def\ee{\end{equation}}

\def\bea{\begin{eqnarray}}
\def\eea{\end{eqnarray}}

\let\a=\alpha

\begin{document}\date{22 January  2018}
\vspace*{-1.0cm}
\hfill DMUS-MP-18-02
\\
\vspace{1.0cm}

\title{A note  on the Hyper--CR equation, and gauged $N=2$ supergravity}
\author{Maciej Dunajski}
\address{Department of Applied Mathematics and Theoretical Physics\\ 
University of Cambridge\\ Wilberforce Road, Cambridge CB3 0WA\\ UK.}
\email{m.dunajski@damtp.cam.ac.uk}
\author{Jan Gutowski}
\address{
Department of Mathematics\\
University of Surrey\\
Guildford, GU2 7XH, UK}
\email{j.gutowski@surrey.ac.uk}
\author{Wafic Sabra}
\address{ Centre for Advanced Mathematical Sciences and Physics Department\\
American University of Beirut, Beirut, Lebanon.} 
\email{ws00@aub.edu.lb}
\begin{abstract} 
We construct a new class of solutions to the dispersionless hyper--CR equation,
and show how any solution to this equation gives rise to a supersymmetric Einstein--Maxwell
cosmological space--time in $(3+1)$--dimensions.
\end{abstract}   
\maketitle
\section{Introduction}
Let $H=H(x, y, t)$ satisfy the partial--differential equation
\be
\label{2nd_order}
H_{xt}-H_{yy}+H_{y}H_{xx}-H_{x}H_{xy}=0.
\ee
This equation is integrable by twistor transform \cite{D03},
and by the Manakov--Santini type inverse--scattering procedure 
\cite{MS, GSW}.
It arises in $(2+1)$--dimensional  Einstein--€"Weyl geometry, where its solutions characterise the Einstein--Weyl spaces of the hyper--CR type. It also appears
in several other integrable constructions \cite{baran, BK, DS5, 
DK14, fer, mich, pav, sergyeyev}.

In this note we shall construct all solutions to (\ref{2nd_order}),  where
the linear and nonlinear terms in (\ref{2nd_order}) vanish separately. 
This will be done in \S\ref{section3}. In \S\ref{section4} we shall show how
solutions to (\ref{2nd_order}) lift to supersymmetric solutions
to $N=2$ pseudo--supergravity in $3+1$ space--time dimensions.
\subsection*{Acknowledgements} 
MD is supported by the STFC consolidated grant  ST/P000681/1.
JG is supported by the STFC consolidated grant ST/L000490/1. 
WS is supported in part by the National Science
Foundation under grant number PHY-1620505.
MD and JG thank the American University of Beirut for hospitality when 
some of this work was undertaken.
\section{Hyper--CR equation}
\label{section3}
Consider three one--forms on a three--dimensional manifold $B$
\[
{\bf e}^1=dx-udy+wdt, \quad {\bf e}^2=dy-udt, \quad {\bf e}^3=dt,
\]
where $(x, y, t)$ is a local coordinate system on $B$, and $u, w$ are two 
functions of $(x, y, t)$. 
Let 
\[
\omega=u_xdy+(uu_x+2u_y)dt, \quad\mbox{and}\quad V=\frac{u_x}{2}
\]
be another one--form, and a function on $B$. The
Gauduchon--Tod
system of equations \cite{GT_paper}
\be
\label{GT}
d{\bf e}^i=\frac{1}{2}{\omega}\wedge {\bf e}^i-V\ast {\bf e}^i, \quad i=1, 2, 3
\ee
holds where $\ast$ is the Hodge operator\footnote{
Note that $\ast {\bf e}^1={\bf e}^1\wedge {\bf e}^2, \ast {\bf e}^2=2{\bf e}^1\wedge {\bf e}^3, \ast 
{\bf e}^3={\bf e}^2\wedge {\bf e}^3$.} of a Lorentzian metric on $B$
\[
h={\bf e}^2\odot {\bf e}^2-4 {\bf e}^1\odot {\bf e}^3
\]
if
the functions $(u, w)$ satisfy a system of integrable equations of hydrodynamic type
(the hyper--CR system)
\be
\label{hyper_CR}
u_t+w_y+uw_x-wu_x=0, \quad u_y+w_x=0.
\ee
Conversely, it has been shown in \cite{D03} that a pair $(h, \omega)$ gives rise to an Einstein--Weyl structure:
there exists a torsion--free connection $D$ such that $D h=\omega\otimes h$ and the symmetrised Ricci tensor of $D$ is proportional to $h$ if and only if
the system (\ref{hyper_CR}) holds. The integrability conditions for  (\ref{GT}) are given by the monopole equation
\[
\ast\Big(dV+\frac{1}{2}V\omega\Big)=\frac{1}{2}d\omega.
\]
This equation holds as a consequence of (\ref{hyper_CR}) -- it becomes a derivative of the first
equation in  (\ref{hyper_CR}).
The Bianchi identity implies that no other integrability conditions arise.
\subsection{Conformal invariance}
Both the Einstein--Weyl  condition, and the Gauduchon--Tod system (\ref{GT}) are conformally invariant if
\be
\label{conformal}
{\bf e}^i\rightarrow e^f {\bf e}^i, \quad \omega\rightarrow \omega+2df, \quad V\rightarrow e^{-f} V.
\ee
In particular, it is possible to fix a conformal gauge such that $V\equiv -2\ell^{-1}$ is a constant 
so that the monopole equation reduces to 
\[
d\omega+2{{\ell}}^{-1}\ast \omega=0.
\] 
In this gauge $\omega$
is divergence-free (so this is the Gauduchon gauge  - note that the converse is not true. There is a residual gauge freedom if the Gauduchon gauge has been fixed which allows for non--constant $V$).
\subsection{The  $\psi$--equation}
Let $\psi$ be any $p$--form of conformal weight $m$, so that $\psi\rightarrow e^{mf}\psi$
under  (\ref{conformal}). The {\em weighted exterior derivative}
\[
D\psi\equiv d\psi-\frac{m}{2}\omega\wedge\psi
\]
is a $(p+1)$--form of weight $m$. 
Let us assume that $\psi$ is a one--form. Using the conformal properties of the Hodge operator we verify that the equation
\be
\label{psi_eq}
D\psi=V\ast\psi
\ee
is conformally invariant as the weight of $V$ is $-1$. In \cite{sabra} this equation
has arisen in a gauge where $V=-2\ell^{-1}$ is a constant, and $m=-1$ where it becomes
\[
d\psi+\frac{1}{2}\omega\wedge\psi=-2\ell^{-1}\ast\psi.
\]
There is a particular solution to this equation given 
by $\psi=c\omega$, where $c$ is a constant.
\subsection{Example. The Heisenberg group}
Let us consider a particular solution of
(\ref{hyper_CR}) given by $u=4\ell^{-1}x, w=0$,  where $\ell$ is a constant.  The resulting Lorentizian Einstein--Weyl structure is defined on the nilpotent Lie group:
\be
\label{heisenberg}
h=(dy+4\ell^{-1}xdt)^2-4dxdt, \quad \omega= 4\ell^{-1}(dy+4\ell^{-1}xdt),
\ee
and $V=2\ell^{-1}$ is a constant. A MAPLE--aided computation
shows that the most general solution to  
(\ref{psi_eq}) which does not depend on $y$ is of the form
\[
\psi=c\omega+d(c+k),
\]
where $c=c(x)$ and $k=k(t)$ are arbitrary functions.
\subsection{Differential constraints}
Let us look for a class of solutions to (\ref{hyper_CR}) where both the linear and nonlinear parts of
the first PDE in (\ref{hyper_CR}) vanish separately. The second equation in (\ref{hyper_CR}) 
can be solved in general to give $u=H_x, w=-H_y$, where $H=H(x, y, t)$ satisfies
(\ref{2nd_order})
and then the special 
constraint resulting from $u_t+w_y=0$ is that $H$ is a solution to the wave equation
on the flat background $dy^2-4dxdt$, and additionally $H_{xx}H_y-H_{xy}H_x=0$. An example is provided
by  the fundamental solution
\be
\label{fundamental}
H=\frac{1}{\sqrt{y^2-4xt}}.
\ee
In general we can establish the following
\begin{prop}
Let $(h, \omega)$ be a hyper-CR Einstein--Weyl structure arising from the equation
(\ref{2nd_order}) such that
\be
\label{constraints}
H_{xx}H_y-H_{xy}H_x=0, \quad\mbox{and}\quad H_{xt}-H_{yy}=0.
\ee
Then there exists a local coordinate system $(p, y, t)$ on  $B$ such that $(h, \omega)$ takes
one of the following three forms
\begin{enumerate}
\item[{\bf Class A}]
\be
\label{classa}
h=(dy+pdt)^2-4\Big(\frac{dp}{p}-\frac{\beta_y}{\beta}(dy+pdt)-\frac{\beta_t}{\beta}dt\Big)dt, \quad
\omega=-p(dy+pdt)+2p\frac{\beta_y}\beta dt,
\ee
where $\beta=\beta(y, t)$ satisfies $\beta_t+\beta_{yy}=0$.
\item[{\bf Class B}]
\be
\label{classb}
h=(dy+pdt)^2+4 F dpdt, \quad\omega=-F^{-1}(dy+pdt),
\ee
where $F=F(p)$ is an arbitrary function.
\item[{\bf Class C}]
\be
h=(dy+pdt)^2+4\Big(2K\frac{dp}{p}-\frac{y}{2t}dy+\frac{1}{t}\Big(\frac{y^2}{4t}+K-\frac{py}{2}\Big)dt\Big)dt, \quad \omega=-\frac{p}{2K}\Big((dy+pdt)-\frac{y}{t}dt\Big),
\label{classc}
\ee
where $K=K(tp^2)$ is an arbitrary function.
\end{enumerate}
\end{prop}
\noindent
{\bf Proof.}
We rewrite the non--linear constraint in (\ref{constraints}) 
as $dH\wedge dH_x\wedge dt=0$, and perform a Legendre
transform $G(p, t, y)=H-px$, where  $H_x=p$ and $x=-G_p$, and the constraint can be solved as
\be
\label{GG_constraint}
G(p, y, t)=A(p, t)+pB(y, t).
\ee
Imposing  the wave equation (the linear constraint in (\ref{constraints})) yields
\be
\label{G_equation}
G_{yp}^2-G_{yy}G_{pp}-G_{pt}=0.
\ee
Substituting (\ref{GG_constraint}) and differentiating the resulting expression with respect to $y$ yields
\be
2B_yB_{yy}-pB_{yyy}A_{pp}-B_{yt}=0.
\label{one_more_eq}
\ee
There are two cases to consider.
\begin{itemize}
\item
If $B_{yyy}\neq 0$ then we can solve (\ref{one_more_eq}) for 
$A_{pp}$, and find
\[
A=\mu(p\ln{(p)}-p)+\kappa(t)p+\rho(t),
\]
where $\mu$ is a constant.
Setting $B=-\mu\ln{\beta(y,t)}-\kappa(t)$ removes $\kappa(t)$ from $G$ and reduces
the equation (\ref{G_equation}) to $\beta_t+\mu\beta_{yy}=0$. Rescaling $t$ and $p$ in the resulting
EW structure can be used to set $\mu=1$. The function $\rho(t)$ does not appear in the EW structure, so
can be set to zero. This yields (\ref{classa}).
\item If $B_{yyy}=0$ then  $B=c_1(t)y^2+c_2(t)y+c_3(t)$
and (\ref{G_equation}) implies that
\[
\dot{c_1}=4{(c_1)}^2.
\]
The classification now branches. If $c_1=0$
then $B$ has to be linear in $y$, and
take the form $B=cy+\gamma(t)$, where $c$ is a constant. The equation
(\ref{G_equation}) becomes
\[
A_{tp}=c^2-\gamma_t,
\]
so that $A=c^2 pt-p\gamma(t)+a(p)+b(t)$, where $a, b$ are some arbitrary functions
of  one variable. Substituting this into $G$ shows that $\gamma(t)$ disappears
from the EW structure, and $b(t)$ can be set to zero. The constant $c$ can also 
be set to zero by an affine transformation of the coordinate $p$ in the EW structure.
This yields (\ref{classb}), where $F=a_{pp}$ is an arbitrary function of $p$.
The nilpotent example (\ref{heisenberg}) belongs to this class, and
corresponds to $F=-\l/4$, where $\l$ is a constant.
 
Next consider the case where $c_1=-1/(4t)$ (the constant of integration in the denominator has been set to zero by shifting $t$). The function $c_3(t)$ can be absorbed into $A(t, p)$,
and  (\ref{G_equation}) gives $c_2(t)=c/t$,  where $c$ is a constant. The resulting equation for $A$ is
\[
tpA_{pp}-2t^2A_{pt}+2c^2=0,
\]
which can be solved in terms of an arbitrary function of
$tp^2$. The constant $c$ can be set to zero by shifting
$y\rightarrow y-2c$. The final expression
for the EW structure takes the form (\ref{classc}).
The solution (\ref{fundamental}) belongs to this class
with $A_p=2^{-4/3}(tp^2)^{-1/3}$.
\end{itemize}
\koniec
\section{Einstein--Maxwell cosmological space--times}
\label{section4}
It is known \cite{GS1, DGST, klemm} that Riemannian solutions to the hyper--CR Einstein--Weyl
equations lift to supersymmetric solutions to the minimal gauged supergravity
in four dimensions (see also \cite{meessen} where the EW geometry appears in supergravity in a rather different context).
Here, following \cite{sabra}, we present an analytic continuation of these constructions where the underlying base Einstein--Weyl manifold is Lorentzian, 
and the resulting four--dimensional theory admits pseudo--supersymmetry.
The bosonic content of the theory consist of a metric, and a one--form which
satisfy the cosmological  Einstein--Maxwell equations with non--standard coupling
between the Maxwell and the Einstein terms.
The metric and the one--form  given by 
\begin{eqnarray*}
g&=&\Big(\frac{\l}{\sin{\a}}d\a-\frac{\l}{2}\cos{\a}\;\omega
+\sqrt{2}{\sin{\a}}\;\psi\Big)^2+\frac{1}{\sin{\a}^2}h\\
A&=&\frac{\sqrt{2}}{2}\sin{2\a}\;\psi-\frac{\l}{4}\cos{2\a}\;\omega
\end{eqnarray*}
satisfy the Einstein--Maxwell equations
\[
R_{ab}+3\l^{-2}g_{ab}+2F_{ac}{F_b}^c-\frac{1}{2}|F|^2g_{ab}=0, \quad d\star_g F=0
\]
iff equations (\ref{GT}) and (\ref{psi_eq}) hold with $V=-2\ell^{-1}$.

Computing the curvature invariants $|F|^2$ and $|\mbox{Riemann}|^2$ suggests
that $\alpha=0, \pi, 2\pi, \dots$ is just a coordinate singularity. A coordinate 
transformation
\[
\sin{\a}=(\cosh{{\l}^{-1} p})^{-1}, \quad \cos{\a}=\tanh{{\l}^{-1} p}
\]
brings the metric and the Maxwell potential to a regular form
\[
g=\Big(dp-\frac{\l}{2}\tanh{(\l^{-1}p)}\;\omega+
\frac{\sqrt{2}}{\cosh{(\l^{-1}p)}}\;\psi\Big)^2+\cosh^2{(\l^{-1}p)}h.
\]
We  can attempt to take a limit where the cosmological  constant vanishes, or equivalently
$\l\rightarrow  \infty$. In this limit 
$\cosh^2{(\l^{-1}p)}\rightarrow 1$, and $\l\tanh{(\l^{-1}p)}\rightarrow p$
so that
\[
g\rightarrow \Big(dp-\frac{1}{2}p\omega\Big)^2+h, \quad F\rightarrow -\frac{\l}{4}d\omega.
\]
The limit exists if $\omega$ depends explicitly on $\l$, and $F$ does not blow up.
This is the case for the Heisenberg group example (\ref{heisenberg}) where $g$ becomes the Minkowski metric, and $F=0$ in the limit.

\end{document}